\documentclass[lettersize,journal]{IEEEtran}
\usepackage{amsmath,amsfonts}
\usepackage{algorithmic}
\usepackage{algorithm}
\usepackage{array}
\usepackage{textcomp}
\usepackage{stfloats}
\usepackage{url}
\usepackage{verbatim}
\usepackage{graphicx}
\usepackage{cite}
\usepackage{multirow}
\usepackage{color}
\usepackage{bm}
\usepackage{hyperref}
\hypersetup{
	colorlinks=true,
	linkcolor=red}
\usepackage{float}
\usepackage{booktabs}
\newcommand{\etal}{\textit{et al.}}
\usepackage{pifont}
\usepackage{subcaption} 
\usepackage{setspace}
\usepackage{fancyhdr}

\begin{document}

\title{Spectral-wise Implicit Neural Representation\\ for Hyperspectral Image Reconstruction}

\author{Huan Chen$^{*}$, Wangcai Zhao$^{*}$, Tingfa Xu$^{\dagger}$, Guokai Shi, Shiyun Zhou, Peifu Liu, Jianan Li$^{\dagger}$
\thanks{}


}


\maketitle
\thispagestyle{fancy}
\lhead{}
\lfoot{}
\cfoot{\small{Copyright © 20xx IEEE. Personal use of this material is permitted. However, permission to use this material for any other purposes must be obtained from the IEEE by sending an email to pubs-permissions@ieee.org.}}

\begin{abstract}
Coded Aperture Snapshot Spectral Imaging (CASSI) reconstruction aims to recover the 3D spatial-spectral signal from 2D measurement. Existing methods for reconstructing Hyperspectral Image (HSI) typically involve learning mappings from a 2D compressed image to a predetermined set of discrete spectral bands. However, this approach overlooks the inherent continuity of the spectral information. In this study, we propose an innovative method called Spectral-wise Implicit Neural Representation (SINR) as a pioneering step toward addressing this limitation. SINR introduces a continuous spectral amplification process for HSI reconstruction, enabling spectral super-resolution with customizable magnification factors. To achieve this, we leverage the concept of implicit neural representation. 
Specifically, our approach introduces a spectral-wise attention mechanism that treats individual channels as distinct tokens, thereby capturing global spectral dependencies.
Additionally, our approach incorporates two components, namely a Fourier coordinate encoder and a spectral scale factor module. The Fourier coordinate encoder enhances the SINR's ability to emphasize high-frequency components, while the spectral scale factor module guides the SINR to adapt to the variable number of spectral channels. 
Notably, the SINR framework enhances the flexibility of CASSI reconstruction by accommodating an unlimited number of spectral bands in the desired output.
Extensive experiments demonstrate that our SINR outperforms baseline methods. By enabling continuous reconstruction within the CASSI framework, we take the initial stride toward integrating implicit neural representation into the field. The code will be released at~\url{https://github.com/chh11/SINR}.
\end{abstract}


\section{Introduction}
\IEEEPARstart{H}{yperspectral} images (HSIs) provide valuable insights into our environment, capturing intricate spatial and continuous spectral information in three-dimensional (3D) tensors. Every pixel's intensity across the spectral range is meticulously documented, forming a comprehensive visual mosaic~\cite{li2022exploring,liu2021global,xu2020hyperspectral,yang2022hyperspectral, wang2021hyperspectral,dopido2012new, xie2020multiscale}. However, this wealth of data presents challenges. Conventional 2D sensors require multiple captures to construct a 3D data cube, limiting real-time scene imaging~\cite{zhang2021deeply}.
The inventive Coded Aperture Snapshot Spectral Imaging (CASSI) system~\cite{zhang2018fast,arce2013compressive, wagadarikar2008single,wang2015high,tang2022single} emerges as a solution. It employs mask modulation and spectral compression to generate a compacted 2D tensor. This tensor is then reconstructed into a complete 3D HSI cube using intricate algorithms~\cite{qiao2020snapshot,meng2021self}.

\begin{figure}[tbp]
\centering
\includegraphics[width=\linewidth]{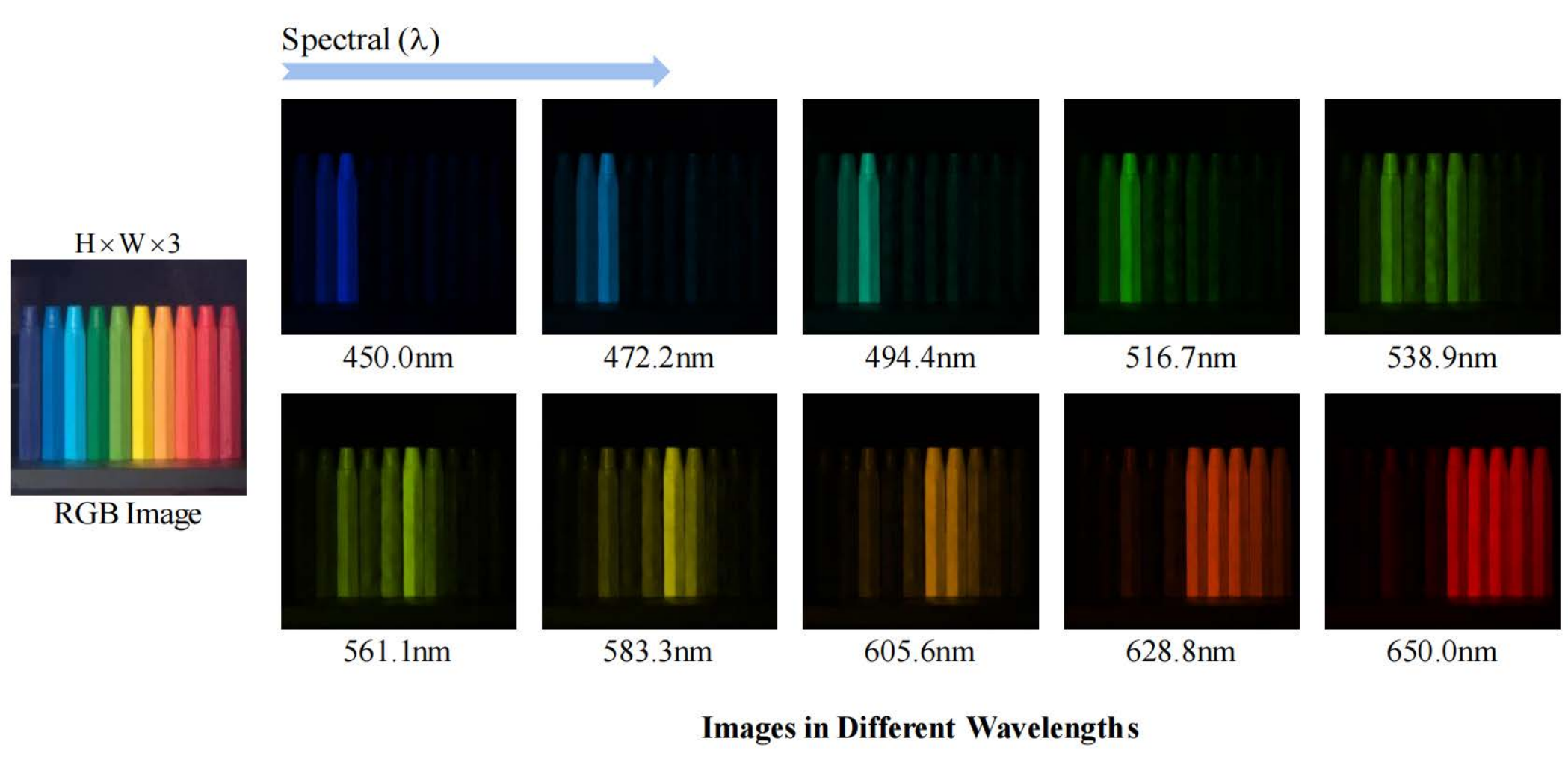}
\caption{Correlation and Complementarity among HSI Spectral Channels
Spatial information in adjacent spectral channels displays significant correlation, as evidenced by the similarity between 2D images rendered at 450.0$\rm {nm}$ and 472.2$\rm {nm}$. Conversely, spatial information with substantial wavelength differences complements each other. This is evident in the corresponding 2D images at 450.0$\rm {nm}$ and 650.0$\rm {nm}$, which capture distinct portions of the RGB image's spatial structure.}
\label{fig1}
\end{figure}

Expanding on CASSI, researchers focus on precise spatial information recovery. Conventional methods use prior regularization like total variation~\cite{kittle2010multiframe,wang2015dual}, non-local similarity~\cite{wang2016adaptive,yang2021transpose}, and sparsity~\cite{wagadarikar2008single,lin2014spatial}. Deep convolutional neural networks map 2D measurements to HSIs. However, most treat HSI spectra as discrete bands~\cite{hu2022hdnet,meng2020snapshot,cai2022coarse}.
In the natural visual world, spectra and sensor responses are continuous. Each HSI pixel holds a comprehensive spectral profile. Models using spectral separation-based methods are limited by fixed spectral band reconstruction. Changing HSI bands requires retraining, elevating time costs and impeding CASSI's practicality.
Discrete HSI reconstruction loses spectral features~\cite{xu2021continuous}, sacrificing fidelity.

This study aims to enhance CASSI reconstruction flexibility and optimize spectral fidelity. We propose a single model capable of reconstructing HSIs with various spectral resolutions, negating the requirement to train separate models for distinct spectral band counts. Our motivation draws from the achievements of Implicit Neural Representation (INR) networks, as seen in their success in 3D object and scene representation~\cite{michalkiewicz2019implicit,chen2019learning}, and continuous 2D image super-resolution~\cite{chen2021learning}.
Hence, we approach the CASSI reconstruction problem through an INR perspective. INR involves establishing relationships between continuous coordinate mappings and domain-specific signals using Depth Implicit Function (DIF)~\cite{xu2021continuous}. The encoder predicts latent codes from various inputs to share representation functions among input instances~\cite{zheng2021deep,jiang2020local}. Equally distributed latent codes can represent a scene across original coordinates. INR achieves continuous output through implicit neural interpolation, which computes a weighted average of predictions from neighboring coordinates~\cite{lee2021meta}. The DIF is parameterized by a decoder, usually a coordinate-based Multi-Layer Perceptron (MLP).

According to the INR concept, two key elements contribute to continuous representation: continuous coordinates and domain-specific signals. Firstly, the natural visual world's spectrum exhibits continuous variation, allowing any individual pixel within an HSI to approximate an all-encompassing spectral profile. This fulfills the prerequisite for continuous coordinates in INR. Secondly, the signal associated with the CASSI spectral reconstruction task corresponds to a continuous 2D spatial signal for each wavelength. Hence, INR can offer a promising avenue to overcome the inherent discontinuity constraints within reconstruction techniques.

However, directly applying the continuous representation of INR to CASSI reconstruction presents a bottleneck. HSIs differ from RGB images in that they exhibit spectral correlation and complementarity. As illustrated in Fig.~\ref{fig1}, the similarity of image signals between adjacent spectral channels is relatively high, and each channel captures a fraction of the same scene. Hence, in the implicit representation of HSIs, adjusting the receptive field adaptively according to the number of reconstructed bands is inadequate if the interpolation of a few fixed points in the vicinity is calculated as in earlier work~\cite{chen2021learning}. Fixed interpolation of neighboring bands primarily captures locally similar information, failing to capture distant dependencies dynamically. This constraint hampers the flexibility of reconstructing arbitrary spectral bands.

Furthermore, prior research~\cite{rahaman2019spectral,cao2019towards} has unveiled that coordinate-based MLPs inherently grasp low-frequency information, struggling to encapsulate high-frequency details, known as spectral bias. This phenomenon also exists in spectral reconstructions, where the absence of high-frequency data characterizing intricate structures results in blurred spectral details.

As a remedy, our work delves into CASSI spectral continuous reconstruction and introduces a novel method called Spectral-wise Implicit Neural Representation (SINR). SINR fully masters the relationship between latent codes and continuous spectral coordinates, exhausting the model's representation to enable the reconstruction of HSIs with arbitrary spectral bands through a unified model.
SINR consists of three components, Spectral-Wise Attention (SWA), Fourier Coordinate Encoder (FCE), and a Reconstructed Head (RH).

Based on the inherent similarities and complementary nature of HSIs, we introduce a novel spectral-wise attention mechanism. This innovation treats each individual spectral channel as a token, facilitating the capture of inter-spectrum dependencies spanning long distances. Unlike traditional fixed local interpolation, our SWA breaks free from confined receptive fields, enabling broader spatial coordinate to latent code interactions.

To address spectral bias, we utilize an FCE to map input coordinates into a high-dimensional Fourier space. This counters the rapid frequency decay in coordinate-based MLPs. By integrating high-dimensional Fourier spectral features~\cite{oechsle2019texture,tancik2020fourier}, we guide the network towards emphasizing intricate high-frequency information.

Furthermore, our method encompasses the reconstruction of a spectral scale factor, which functions as an amplification factor. 
These layouts enhance each other's success and the quality of the reconstructed image.

The entire network is end-to-end trainable. Extensive experiments verify that the proposed method is a promising solution over the baseline by a large margin in both comprehensive quantitative metrics and perceptual quality. In addition, our SINR outperforms state-of-the-art methods in terms of spectral fidelity. This noteworthy achievement holds true even when compared against models specifically trained for a particular band count. 

SINR achieves unprecedented arbitrary spectral magnification in the CASSI reconstruction task, surpassing existing methods that rely on separate spectral bands for reconstruction. 
This eliminates the need for retraining models for each number of spectral bands reconstructed, which has significantly limited the application of CASSI in the past. Through continuous spectral band reconstruction, SINR empowers the training of a model capable of reconstructing any conceivable number of spectral bands. This pivotal advancement dramatically enhances the inherent flexibility of the CASSI system. 

Our main contributions are summarized as follows:  
\begin{itemize}
	\item We are the first to introduce implicit neural representations into the HSI reconstruction of the CASSI system, accomplishing continuous spectrum reconstruction, which makes it possible to reconstruct HSIs for an arbitrary number of spectral bands by training only one model.
	\item We propose novel spectral-wise attention to capturing inter-spectral long-range relationships along the spectral dimension, further improving reconstruction.
	\item We incorporate structural designs such as Fourier coordinate encoder and scale factor into the model. These components improve the adaptability of the implicit function to various scales and frequency information.
\end{itemize}

\section{Related Work}
\subsection{Implicit Neural Representation}
In the natural physical world, objects exhibit a continuous or nearly continuous nature. However, conventional research has predominantly described objects using discrete representations such as point clouds, voxels, mesh grids, and pixels. To transcend this limitation, the concept of Implicit Neural Representation (INR) emerged. INR harnesses continuously differentiable deep implicit functions to map coordinates onto processed signals. This approach enables the emulation of the continuous essence found in natural objects, facilitating the precise capture of intricate details by skillfully adjusting parameters.

Initially, this concept found extensive utility in 3D scene modeling and object representation~\cite{genova2020local,jiang2020local}. Mildenhall \etal~\cite{mildenhall2021nerf} proposed a neural radiance field that synthesizes complex scene views from sparse input views. Building on the triumph of INR in the 3D realm, numerous studies explore the application of shared implicit spaces for images to establish a universal model for various objects~\cite{sitzmann2020implicit}. Chen~\etal~\cite{chen2021learning} introduced LIIF for representing images of arbitrary resolutions. Xu~\etal~\cite{xu2021continuous} addressed Hyperspectral Imaging (HSI) reconstruction from RGB images by embracing INR. However, INR has not yet been effectively applied to establish a continuous spectrum for Coded Aperture Snapshot Spectral Imaging (CASSI) reconstruction. Given the distinctive attributes of HSIs, we embark on an adaptive exploration of implicit neural functions, striving to provide a tailored solution for continuous representation in HSI reconstruction.

\subsection{Hyperspectral Image Reconstruction}
Our work centers around the CASSI system's 2D compressed signal reconstruction task. The input entails a compressed signal that undergoes modulation via a coded aperture and a prism, while the output yields a reconstructed 3D HSI in which the spatial scale remains unchanged. 
In recent years, deep learning has shown remarkable achievements in image transformation. Various solvers have been introduced in this field. For instance, $\lambda$-Net~\cite{miao2019net}, DGSMP~\cite{huang2021deep}, and TSA-Net~\cite{wang2021tsa} have integrated U-Net with other modules, resulting in commendable accomplishments. However, these reconstruction techniques are all predicated on separating spectral information. In contrast, our objective is to surpass these methods and achieve spectral continuity in the reconstruction process. It is worth noting that all these reconstruction methods are based on the premise of spectral separation, and we aim to realize spectral continuity-based reconstruction within the framework of these existing techniques.

\subsection{Spectral Attention}
Following the original proposal in~\cite{vaswani2017attention}, numerous studies have demonstrated the superiority of the attention mechanism in the vision domain~\cite{cao2021swin,dosovitskiy2020image,ji2019video}. Using the attention mechanism to extract spatial information and applying attention solely to spectral channels remains a difficult challenge. Prior researches~\cite{chen2021learning,xu2021ultrasr,mildenhall2021nerf, hao2022attention} show that feature expansion is primarily accomplished by accessing data from $\rm N$ nearby neighboring latent codes to enhance the information present in each latent code.
This operation requires concatenating the data from each domain point, which results in an N-fold increase in data volume. Furthermore, accessing near-neighbor latent codes restricts the receptive fields to a specific region. We propose a self-attentive mechanism to make long-range feature unfolding in the spectral dimension possible by utilizing the ability to record non-local connections.

\subsection{Spectral Bias}
The spectral bias problem occurs when coordinate-based MLPs learn low-frequency information but neglect high-frequency features. This is due to the rapid frequency falloff in MLPs, as explained by the Neural Tangent Kernel theory~\cite{jacot2018neural}. To address this, Basri and Rahaman~\etal~\cite{basri2020frequency, rahaman2019spectral} proposed using Fourier mapping to overcome spectrum distortion and enable better learning of high-frequency information. Recent research~\cite{mildenhall2021nerf,xu2021ultrasr} in image hyper-segmentation proves that proper spatial coding can improve network performance and reduce spectrum distortion. All these studies point to the same conclusion: Fourier mapping to preprocess the input to coordinate-based MLP helps capture high-frequency information and lift performance. Inspired by the above, this work applies Fourier coding to the continuous coordinates to enhance complex expression.

\section{Method}

\begin{figure}[tbp]
\centering
\includegraphics[width=\linewidth]{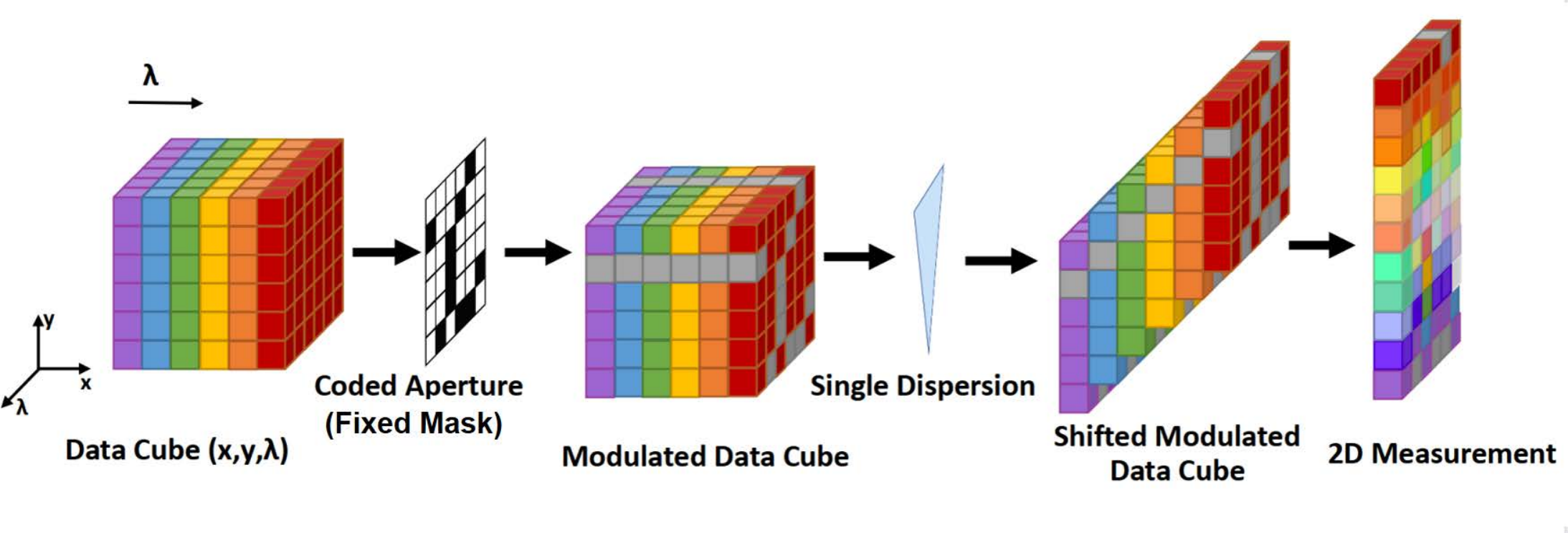}
\caption{The flowchart of CASSI system. Other optical parts are eliminated in favor of a 2D-coded aperture and a dispersion of the CASSI system. For synthetic HSI cubes, measurements can be mathematically computed by modeling the CASSI system.}
\label{fig2}
\end{figure}

\subsection{Preliminaries}
\noindent {\bf{CASSI Imaging System.}} 
The concept of the CASSI system can be seen in Fig.~\ref{fig2}. We denote a 3D HSI cube as $\boldsymbol{F} \in \mathbb{R}^{{ \rm H}\times { \rm W} \times { \rm N_\lambda}}$, where ${ \rm H}$, ${ \rm W}$, and ${\rm N_\lambda}$ represent the height, width, and the number of spectral channels in the HSI cube, respectively. Initially, the HSI cube $\boldsymbol{F}$ undergoes modulation by a physical mask, referred to as the coded aperture. This modulation is achieved through a mask denoted as $\boldsymbol{M} \in \mathbb{R}^{{ \rm H} \times { \rm W}}$:

\begin{equation}
\begin{aligned}
\label{eq:1}
    {\boldsymbol F^{'}}(:,:,n_\lambda) = {\boldsymbol F}(:,:,n_\lambda) \odot  {\boldsymbol M},
\end{aligned}
\end{equation}
where ${\boldsymbol F^{'}}$ represents the modulated HSI, where $n_{\lambda} \in {1, . . ., {\rm N_\lambda}}$ indexes the spectral channels, and $\odot$ signifies element-wise multiplication. After passing through the disperser, ${\boldsymbol F^{'}}$ undergoes a tilt and is sheared along the $y$-axis. The resulting tilted HSI cube is denoted as ${\boldsymbol F^{''}} \in \mathbb{R}^{{ \rm H} \times ({ \rm W}+{ \rm d}({ \rm N_\lambda}-1))\times { \rm N_\lambda}}$, where ${ \rm d}$ denotes the shifting step. The reference wavelength is represented as $\lambda_c$. Consequently, we obtain:
\begin{equation}
\begin{aligned}
\label{eq:2}
    {\boldsymbol F^{''}}(u,v,n_{\lambda}) =  {\boldsymbol F^{'}}(x,y+{ \rm d}(\lambda_n -\lambda_c),n_\lambda),
\end{aligned}
\end{equation}
where $(u,v)$ represents the coordinate system on the detector plane, while $\lambda_n$ signifies the wavelength of the $n_{\lambda}$-th channel. ${\rm d}(\lambda_n - \lambda_c)$ denotes the spatial shift for the $n_{\lambda}$-th channel on ${\boldsymbol F^{''}}$. It's important to note that all the light within the wavelength range $[\lambda_{\rm{min}}, \lambda_{\rm{max}}]$ on the detector is captured, resulting in a continuous signal just as observed in nature. The compressed measurement on the detector, denoted as $y(u,v)$, can be modeled as:
\begin{equation}
\begin{aligned}
\label{eq:3}
    {y}(u,v) = \bm\int_{\rm{min}}^{\rm{max}} {\boldsymbol f}^{''}(u,v,n_{\lambda}){{ \rm d}}\lambda,
\end{aligned}
\end{equation}
where ${\boldsymbol f}^{''}$ is a continuous representation of $\boldsymbol{F^{''}}$. Most reconstructions discretize it, and the captured 2D compressed measurement $\boldsymbol{Y}$ can be further described by:
\begin{equation}
\begin{aligned}
\label{eq:4}
    \boldsymbol{Y} = \bm\sum_{n_{\lambda}=1}^{{\rm N_\lambda}} {\boldsymbol F}^{''}(:,:,n_{\lambda}),
\end{aligned}
\end{equation}
where $\boldsymbol{Y}$ is the 2D measurement of a continuous HSI cube after CASSI imaging.

\noindent {\bf{Image Reconstruction and Implicit Representation.}} 
The initialized input for the reconstruction model ${\boldsymbol{F}}_Y[:,:,n_{\lambda}] \in \mathbb{R}^{{ \rm H} \times { \rm W} \times { \rm N_\lambda}}$ is obtained by reversing the dispersion from the 2D measurement:
\begin{equation}
\begin{aligned}
\label{eq:6}
    \boldsymbol{F}_Y[:,:,n_{\lambda}]: = {\rm{shift}}({\boldsymbol{M}}_{n_\lambda} \odot {\boldsymbol{Y}}),
\end{aligned}
\end{equation}
where ${\boldsymbol{M}}_{n_\lambda}$ is obtained by replicating 2D mask $\boldsymbol{M}$ along the spectral dimension. Then feeding ${\boldsymbol{F}}_Y$ into network:
\begin{equation}
\begin{aligned}
\label{eq:7}
   {\boldsymbol{Z}} = {\boldsymbol E}_{\varphi}(\boldsymbol{F}_Y),
\end{aligned}
\end{equation}
where ${\boldsymbol E}_{\varphi}$, the encoder, represents the adopted reconstruction network learning how to fit perfectly offset compression functions. Acquired ${\boldsymbol{Z}} \in \mathbb{R}^{{ \rm H}\times { \rm W} \times { \rm D} \times { \rm C}}$, whose ${ \rm C}$ correspond to the number of feature dimensions and ${ \rm D}$ is the number of spectral channels of the image.
During the execution of the continuous representation, in keeping with~\cite{chen2021learning}, $\boldsymbol {Z}$ represents latent code of $\boldsymbol I_\text{LR}$ with the original low spectral resolution. The HR image $\boldsymbol I_\text{HR}$ is defined as the image of the target spectrum with high resolution computed by the implicit function. Between $\boldsymbol I_\text{HR}$ and $\boldsymbol I_\text{LR}$ can be linked by the following implicit function:
\begin{align} \label{eq:9}
    { \boldsymbol f}_{\theta}(z,\cdot) : {\boldsymbol X} \rightarrow {\boldsymbol S},
\end{align}
where $\boldsymbol{S} \in \mathbb{R}^{{ \rm H}\times { \rm W} \times { \rm D}} $ is the reconstruction's target values, $\boldsymbol{X} \in \mathbb{R}^{{ \rm H}\times { \rm W} \times { \rm D}} $ denotes spectral coordinates. $z \in \boldsymbol{Z}$ is the reference feature vector in latent space, extracted from $\boldsymbol I_\text{LR}$ by the encoder. ${ \boldsymbol f}_{\theta}(\cdot)$ is the decoding function learned by an MLP with learnable parameters $\theta$.

As previously stated, the spectrum compressed by the CASSI system is a continuous signal. When the provided coordinates are similarly continuous, obtaining a continuous spectrum expression via an implicit function is theoretically conceivable. The upsampled continuous image ${\boldsymbol I_\text{HR}}$ is further defined as:

\begin{align} \label{eq:10}
   {\boldsymbol I}_\text{HR}={ \boldsymbol f}_{\theta}(\boldsymbol X, \boldsymbol Z).
\end{align}

While the INR has proven successful in achieving continuous representation in the spatial domain of RGB images, further advancements are necessary to extend this success to the reconstruction of HSIs, both spatially and spectrally. In RGB images, the spatial information of a pixel is primarily influenced by its immediate neighbors. However, HSIs require the incorporation of multiple surrounding spectral bands to restore the original scene accurately. Merely applying the same approach used for local area reconstruction of specific wavelength images would lead to the loss of information in the reconstructed scene.

Furthermore, implicit neural functions have a tendency to emphasize the learning of low-frequency information, leaving the high-frequency information responsible for capturing spectral intricacies insufficiently learned, thus resulting in blurriness. In summary, the core challenge is to tailor an implicit spectral representation function specifically for HSI reconstruction.

\subsection{Model Overview}

\begin{figure*}[htbp]
\centering
\includegraphics[width=\textwidth]{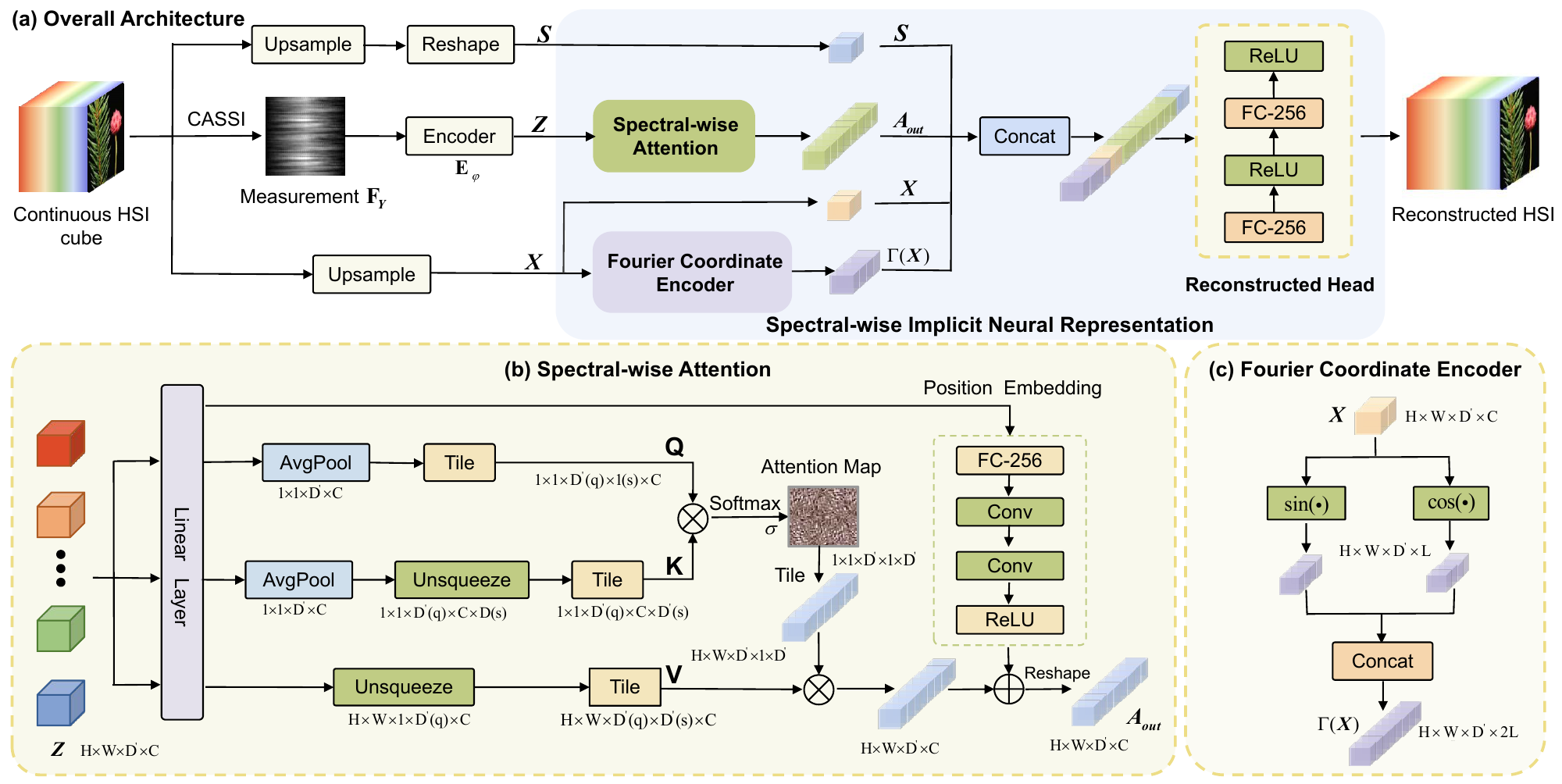}
\caption{Overall network architecture of continuous spectral reconstruction, which comprises an encoder and our Spectral-wise Implicit Neural Representation (SINR). Inputs of SINR are as follows: scale factor $\boldsymbol S$, feature map extracted from an encoder $\boldsymbol Z$,  and spectral coordinates $\boldsymbol X$. }
\label{fig3}
\end{figure*}
 
\begin{figure}[h!t]
\centering
\includegraphics[width=\linewidth]{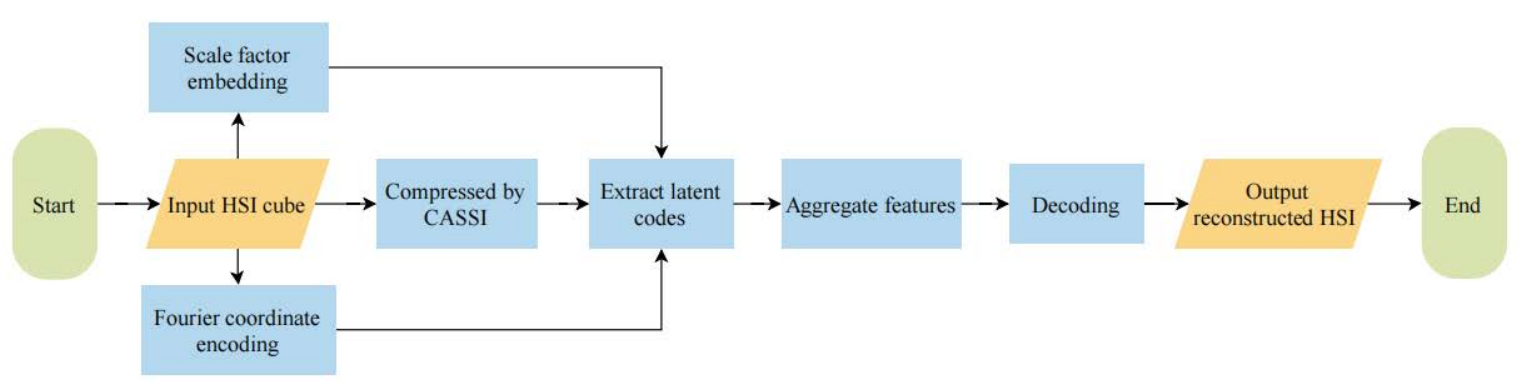}
\caption{The flowchart of the proposed method SINR.}
\label{flowchat}
\end{figure}

The architecture for achieving spectral continuous reconstruction is depicted in Fig.~\ref{fig3} and Fig.~\ref{flowchat} . This approach employs INR to establish a continuous spectrum for the purpose of HSI reconstruction. Given a compressed measurement obtained through the CASSI system, the objective is to reconstruct a high-spectral-resolution image $\hat{\boldsymbol I}\in \mathbb{R}^{{ \rm H}\times { \rm W}\times { \rm D}^{'}}$ from the original limited spectral data ${\boldsymbol I}_\text{LR} \in \mathbb{R}^{{ \rm H}\times { \rm W}\times { \rm D}}$, while accommodating an arbitrary scale factor $r$ that satisfies ${ \rm D}^{'}=r \times { \rm D}$:

\begin{align}
\label{eq:11}
    { \boldsymbol f}_{\theta}(z,\cdot) : {\boldsymbol I}_\text{LR} \rightarrow {\hat{\boldsymbol I}}.
\end{align}

Our Spectral-wise Implicit Neural Representation model (SINR) mainly comprises four parts: i) Spectral-Wise Attention (SWA) which retrieves latent codes corresponding to consecutive spectral coordinates; ii) Fourier Coordinate Encoder (FCE) that maps the coordinates into the high spectral resolution periodically; iii) a Scale Factor (SF) module that delivers scale information to facilitate the precise capture of fine-grained structures; iv) a Reconstructed Head (RH) provides a way for parameterizing the INR by an MLP.

\subsection{Encoder Network}
Given measurement ${\boldsymbol F}_Y$ captured by the CASSI system, the encoder learns a mapping function expressed in a vector space:
\begin{equation}
\begin{aligned}
\label{eq:12}
    {\boldsymbol E}_{\varphi}(\cdot):{\boldsymbol F}_Y \rightarrow {\boldsymbol {F^{''}}}.
\end{aligned}
\end{equation}

To achieve continuous spectral enhancement while maintaining spatial resolution, we utilize spatial-spectral invariance and high-performance encoders. Specifically, we employ the SSI-ResU-Net~\cite{wang2021new} and MST~\cite{cai2022mask} architectures as encoders for extracting latent codes from low-resolution images. An additional output convolution layer is integrated while preserving the original structure. This yields a feature map of dimensions $\mathbb{R}^{{\rm H}\times {\rm W}\times {\rm D}\times {\rm C}}$, with ${\rm C}$ representing the number of latent codes having a shape of ${\rm H}\times {\rm W}\times {\rm D}$. These latent codes are evenly distributed along the spectral dimension. Each code $z$ within $\boldsymbol Z$ signifies spectral information across a continuous region of hyperspectral images, playing a pivotal role in predicting the surrounding signal.

\subsection{Spectral-wise Implicit Neural Representation}
The encoder module plays a crucial role in capturing latent codes. To establish this relationship between continuous coordinates and latent codes, a comprehensive analysis is conducted. The first component of SINR is SWA which treats each channel as a token and enables the gathering of global spectral dependencies. The FCE employs heuristic sinusoidal mappings to translate input coordinates into higher dimensions, guiding the MLP to learn continuous coordinate data in the low-frequency domain. Furthermore, an SF is introduced as global amplification information. Finally, the RH parameterizes the implicit function. The SWA serves as the core component of SINR, while the FCE and SF operate in a plug-and-play manner, enhancing the network's capability to recover fine-grained spectral structures.

\noindent{\bf{Spectral-wise Attention.}} The self-attention mechanism possesses the capacity to capture extensive dependencies among vectors. Thus, our approach utilizes a transformer-based architecture to effectively handle the task of capturing inter-channel spectral features. This choice enables the incorporation of global associations between spectral features and coordinates, integrating contextual information from all spectral bands. Each spectral feature map is treated as a token, and self-attention is applied along the spectral dimension to capture the dependencies and relationships among these spectral features.

The encoder's output, represented as feature map $\boldsymbol Z$, serves as the input for SWA. This feature map, denoted as ${\boldsymbol Z}_\text{HR}$, is interpolated into $\mathbb{R}^{{ \rm H}\times { \rm W}\times { \rm D}^{'}\times { \rm C}}$. A linear layer generates the attention map for the spectral axis, encompassing query $\boldsymbol Q$, key $\boldsymbol K$, and value $\boldsymbol V$, all maintaining the full spectral resolution. Given the sparse distribution of spatial information across the spectrum, average pooling aggregates the spatial details of $\boldsymbol Q$ and $\boldsymbol K$, enhancing computational efficiency. The spatial information of $\boldsymbol V$ remains constant to prevent undue loss of spatial details. Subsequently, $\boldsymbol Q$, $\boldsymbol K$, and $\boldsymbol V$ undergo dimensional expansion and replication, yielding query $\boldsymbol Q \in \mathbb{R}^{1 \times 1 \times { \rm D}^{'}\times { \rm D}^{'}\times { \rm C}}$, key $\boldsymbol K \in \mathbb{R}^{1 \times 1 \times { \rm D}^{'}\times 1\times { \rm C}}$, and value $\boldsymbol V \in \mathbb{R}^{{ \rm H}\times { \rm W}\times { \rm D}^{'}\times { \rm C}}$. This processing of the input can be summarized as follows:

\begin{equation}
\begin{aligned}
\label{eq:13}
    {\boldsymbol Q} &= \boldsymbol f_\text{R1}(\boldsymbol f_\text{AVP}({\boldsymbol Z}_\text{HR}))) \\  
    {\boldsymbol K} &=  \boldsymbol f_\text{R2}( \boldsymbol f_\text{AVP}( {\boldsymbol Z}_\text{HR})))\\
    {\boldsymbol V} &=  \boldsymbol f_\text{R3}({\boldsymbol Z}_\text{HR})),
\end{aligned}
\end{equation}
where $\boldsymbol{f}_\text{R1}(\cdot)$, $\boldsymbol{f}_\text{R2}(\cdot)$, $\boldsymbol{f}_\text{R3}(\cdot)$ represent the unsqueeze and tile operations,
$\boldsymbol f_\text{AVP}(\cdot)$ denotes 3D adaptive average pooling operation. ${\boldsymbol Q}$ stands for the query channel in the upsampled image, and ${\boldsymbol K}$ is the key point in the input LR image. 
Moreover, ${\boldsymbol V}$ is the corresponding feature map where both ${\boldsymbol Q}$ and ${\boldsymbol K}$ can be viewed as continuous so that the continuity is reserved when remapping to the value vector ${\boldsymbol V}$:

\begin{align}\label{eq:14}
{\boldsymbol A} &= {\rm Softmax}(\sigma {\boldsymbol Q}{\boldsymbol K}^\top) {\boldsymbol V}.  
\end{align}

The relationship between spectral channels and coordinates is captured using the attention map ${\boldsymbol A}$, calculated between ${\boldsymbol K}$ and ${\boldsymbol Q}$. A learnable parameter $\sigma$ is fine-tuned adaptively to address substantial spectral density variations across channels, while ${\boldsymbol K}^\top$ represents the transpose of ${\boldsymbol K}$. The dot product operation computes the query with all keys, and the $\rm {Softmax}$ function determines the weights for the values. This process yields the global spectral data, projected onto ${\boldsymbol Z}_\text{HR}$.

To incorporate positional information, we introduce position encoding to the SWA. Reflecting the spectral band ordering in the HSI spectrum, we embed the relative position details of spectral bands into the attention outcomes. This involves the use of the position encoding function $\boldsymbol f_\text{pos}(\cdot)$, which includes a linear layer, two $3\times 3 \times1$ convolution layers, a ReLU activation function, and reshape operations. This function is applied to ${\boldsymbol V}$ and can be expressed as:
\begin{align}\label{eq:15}
   {\boldsymbol A}_\text{out} &= {\boldsymbol A} + \boldsymbol f_\text{pos}({\boldsymbol V}).
\end{align}

Consequently, the output of SWA, denoted as ${\boldsymbol A}_\text{out} \in \mathbb{R}^{{\rm H} \times {\rm W} \times {\rm D}^{'} \times {\rm C}}$, is obtained.

\noindent {\bf{Fourier Coordinate Encoder.}} To tackle spectral bias and optimize the utilization of high-frequency details during reconstruction, we apply Fourier mapping to elevate the 1D spectral coordinates into higher dimensions prior to inputting them into the Reconstructed Head. This approach counters the MLP's inclination towards specific frequency ranges and minimizes structural distortion. We leverage the NTK concept to render the mapping stationary or shift-invariant, thereby regulating the spectrum of frequencies learnable by the respective MLP~\cite{tancik2020fourier}. The mapping function is denoted as $\boldsymbol{\Gamma}(\cdot)$:
\begin{equation}
\begin{aligned}
\label{eq:16}
    \boldsymbol{\Gamma}(x) &= [ {\rm cos}({\omega_i}x), {\rm sin}({\omega_i}x)]^\top,
\end{aligned}
\end{equation}
where the initial values for the frequency parameters $\omega_i$ of the coordinates $x$ are set as $2e^i$, where $i \in \{1, 2,..., \rm L\}$, and these parameter frequencies are refined during subsequent training. The encoding functions $\rm{sin(\cdot)}$ and $\rm{cos(\cdot)}$ are heuristic sinusoidal mappings applied to the coordinates of various frequency functions, independently paired for encoding. This conversion transforms the space from $\mathbb{R}^1$ to $\mathbb{R}^{2\rm L}$. Following the FCE, the predictions of SINR, denoted as $\hat{\boldsymbol I}$, can be formulated as:

\begin{equation}
\begin{aligned} \label{eq:17}
    {\hat{\boldsymbol I}} &= \boldsymbol f_\theta ({\boldsymbol A_\text{out}},(\boldsymbol X,{\boldsymbol{\Gamma}}({\boldsymbol X}))).
\end{aligned}
\end{equation}
Additionally, we combine the original coordinates $\boldsymbol X$ with the embedded coordinates to further bolster model stability and information propagation. The influence of the mapping dimensions is examined in Section~\ref{sec:ablation}.

\noindent {\bf{Scale Factor.}} 
The integration of global class token information has been shown to be pivotal for effective model training, as exemplified by the Vision Transformer architecture~\cite{dosovitskiy2020image}. Building on this idea, we fuse the coordinate information—used as a reference for image shape prediction—with the scale information of the high-resolution image. The scale factor conveys global magnification details. Equation~\ref{eq:17} is reformulated as:
\begin{equation}
\begin{aligned}\label{eq:18}
    {\hat{\boldsymbol I}} &= \boldsymbol f_\theta ({\boldsymbol A}_\text{out},{\boldsymbol X},{\boldsymbol{\Gamma}}({\boldsymbol X}),{\boldsymbol S}).
\end{aligned}   
\end{equation}

\noindent {\bf{Reconstructed Head.}} 
We concatenate the query coordinates ${\boldsymbol X}$ and ${\boldsymbol{\Gamma}}({\boldsymbol X})$, along with the feature vector ${\boldsymbol A}_\text{out}$ and the scale factor ${\boldsymbol S}$ from Equation~\ref{eq:18}, along the channel dimension. This concatenated input is fed into the Reconstructed head, which parameterizes the implicit neural function and produces the predicted intensity $\hat{\boldsymbol I}$.
The function $\boldsymbol f_\theta(\cdot)$ is shared across all images and is implemented using an MLP.
The structure of the MLP is illustrated in the bottom right corner of Fig.~\ref{fig3}, consisting of two fully connected layers with 256 neurons each, separated by a ReLU activation.

\subsection{Optimization}
Our model is exclusively trained at the ($\times 1$) scale, corresponding to cases where input and output image spectral bands are identical. Nonetheless, to assess its generalization capability, we evaluate the SINR's performance on out-of-scale spectral bands, whose scale is unknown. For this purpose, we interpolate the original input image's spectral bands into ${ \rm D}$ bands and extract a spatial region of ${ \rm H}\times { \rm W}$ size. After applying rotation and flipping, we employ this augmented data as encoder input. During training pair creation, we randomly select ${ \rm H}\times { \rm W}\times { \rm D}$ pixels from a high-resolution patch as the ground truth.

The objective of the training is to minimize the differences between the reconstructed HSIs and the ground truth. We utilize the standard $\boldsymbol{L}_1$ distance as the loss function for network optimization:

\begin{equation}
\begin{aligned} \label{eq:19}
    \mathcal{L} = \frac{1}{\rm N} \sum_{i}^{\rm N} {\vert \hat{\boldsymbol I}-\boldsymbol I\vert},
\end{aligned}
\end{equation}

\noindent where $\rm N$ is the total number of sampled bands, $x_i$ is the coordinate of any sampled pixel, $\boldsymbol I$ is the ground-truth pixel value at the high spectral resolution, and $\hat{\boldsymbol I}$ is the predicted pixel value by SINR. For inference, we query the coordinates of all spectral bands in the target domain to recover the full up-sampled image.

\section{Simulation Experiments}
\begin{figure*}[h!t]
\centering
\includegraphics[width=\textwidth]{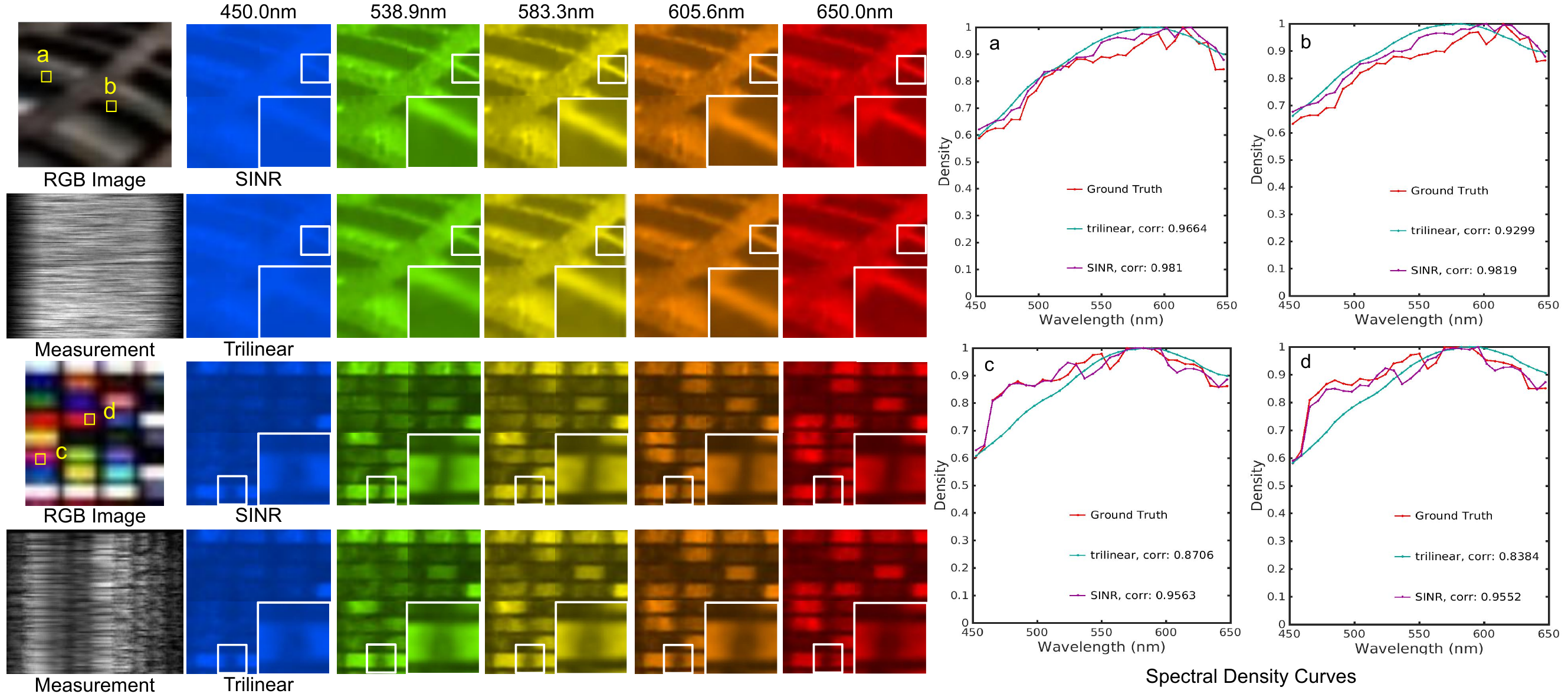}

\caption{Comparisons of reconstruction results on ICVL. SINR enables the reconstructed image to have sharper edges. The right part is the relative spectrum of points a, b, c, and d. The reconstruction spectrum of our method (the purple line) is closer to the ground truth (the red line). Corr refers to the correlation coefficient between the spectral curve of the reconstructed image and the ground truth.}
\label{fig4}
\end{figure*}

\subsection{Experimental Setups}
\noindent {\bf{Dataset.}} We conducted experiments using the ICVL~\cite{arad2016sparse}, Harvard~\cite{arad2016sparse}, KAIST~\cite{choi2017high}, and CAVE dataset~\cite{park2007multispectral}. 

The ICVL dataset includes 201 spectral images, each with 260 bands. We selected 241 bands from 450$\rm {nm}$ to 650$\rm {nm}$ for our experiments. To prevent overfitting, we excluded similar scenes and subjects, ending up with 84 images for training and 10 for testing. These test images were cropped into 4761 smaller $60\times60$ pixel images.

The Harvard dataset~\cite{chakrabarti2011statistics} comprises 50 HSIs from the CRI Nuance FX spectral camera. Each image has 31 spectral bands spanning 420$\rm {nm}$ to 720$\rm {nm}$ and a resolution of $1024\times1392$ pixels. 41 images are used for training and the remaining ones are used for testing, setting the patch size to $120\times120$ pixels.

We also conducted experiments on the KAIST and CAVE datasets, following MST~\cite{cai2022mask}. The KAIST dataset holds 10 HSIs, while the CAVE dataset contains 205 images sized $1024\times1024$ pixels. We selected 28 bands for both from 450$\rm {nm}$ to 650$\rm {nm}$. The CAVE dataset was used for training, and the KAIST dataset for testing, focusing on a $256\times256$ pixel region at the mask center for simulation.\\

\noindent {\bf{Implementation Details.}}
In the CASSI system, we set the shift step $\rm d$ for the dispersion tilt data cube and the reverse dispersion to 2. We utilize the Adam optimizer with an initial learning rate of $4\times10^{-4}$ and a batch size of 4. The FCE dimension $\rm L$ remains fixed at 12 across all experiments, except for the ablation study where we explore the impact of mapping dimension on the model.

To comprehensively evaluate the performance of SINR, we employ a diverse set of metrics, encompassing both spatial and spectral aspects. These metrics include Peak-Signal-to-Noise Ratio (PSNR), Structural Similarity Index Measure (SSIM), Spectral Angle Mapping (SAM), Visual Signal-to-Noise Ratio (VSNR)~\cite{chandler2007vsnr}, Weighted Signal-to-Noise Ratio (WSNR)~\cite{zhou2020weighted}, Noise Quality Measure (NQM~\cite{841940}), Information Fidelity Criterion (IFC~\cite{1532311}), Universal Quality Index (UQI~\cite{995823}), and Visual Information Fidelity (VIF~\cite{1576816}). 

PSNR benchmarks overall image quality via pixel-level differences. SSIM emphasizes structural similarity and perception. VSNR and VIF highlight visual perception with human visual traits, while WSNR captures frequency component significance. IFC focuses on high-frequency traits. NQM assesses noise quality and insights in reconstructed images. SAM quantifies spectral accuracy in reconstruction. UQI comprehensively evaluates spatial and spectral features via multiple factors.

\begin{figure*}[h!t]
\centering
\includegraphics[width=\textwidth]{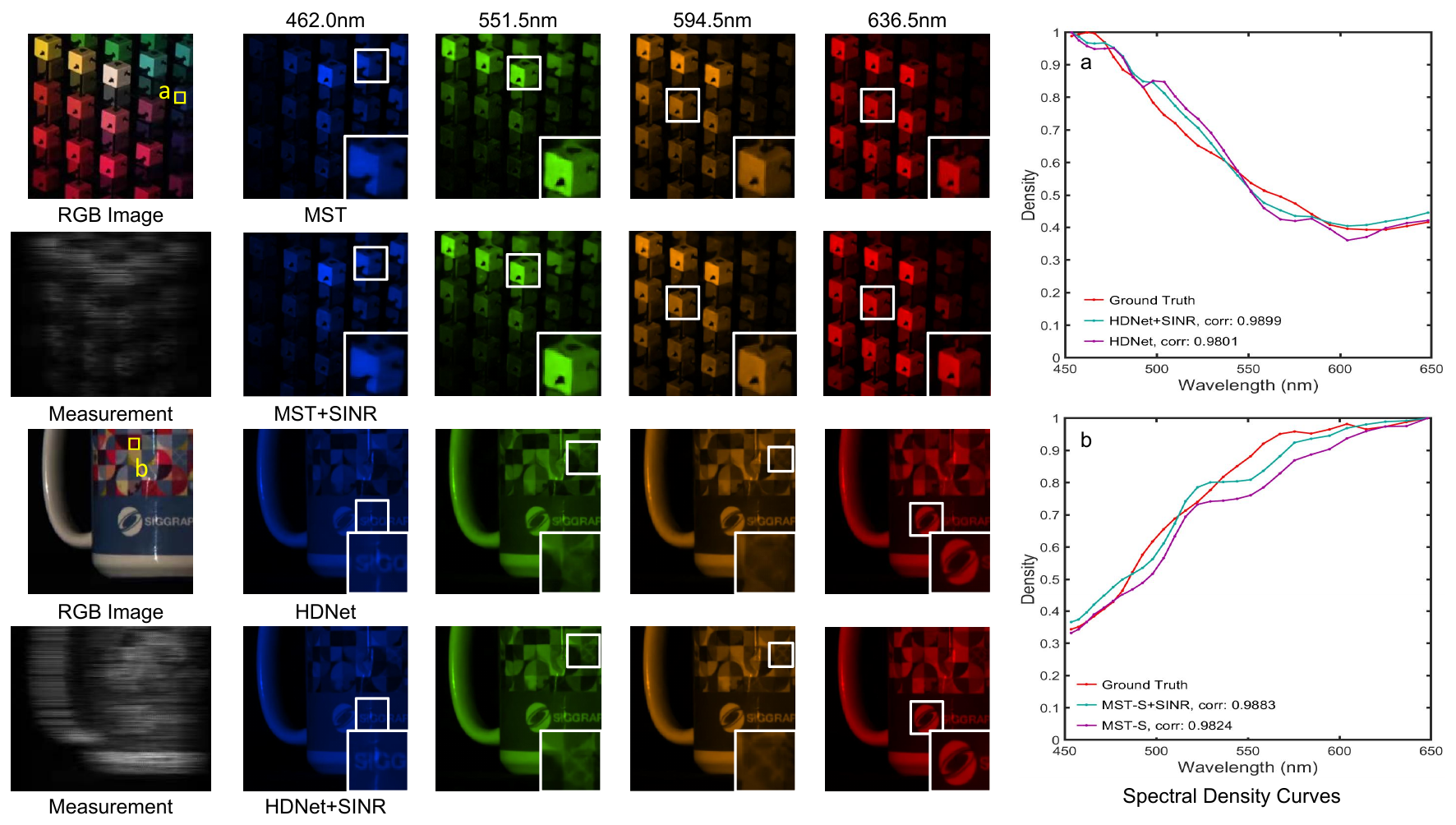}
\caption{Simulated HSI reconstruction comparison of a randomly selected channel on the CAVE and KAIST datasets. SINR reconstructs a clearer and sharper fine structure. The right part is the relative spectrum of two points. The reconstruction spectrum of the SINR (the green line) is closest to the ground truth (the red line).}
\label{fig5}
\end{figure*}
\begin{figure*}[h!t]
\centering
\includegraphics[width=\textwidth]{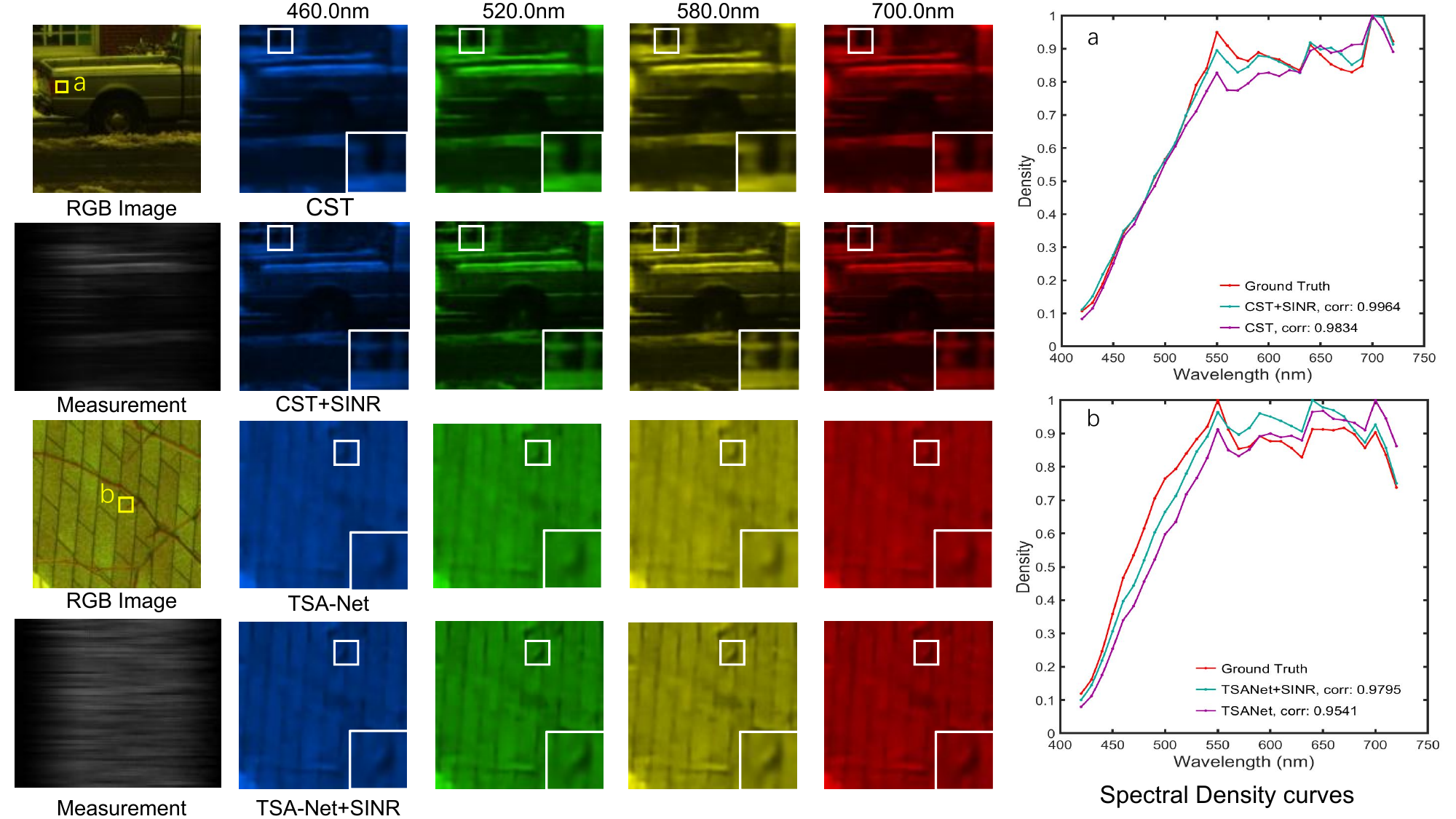}
\caption{Comparison of simulated HSI reconstruction on selected channels from the Harvard dataset reveals that the SINR method yields superior results. The reconstruction spectra of SINR (represented by the green line) closely match the ground truth (represented by the red line).}
\label{harvad}
\end{figure*}

\subsection{Quantitative Comparisons} 

\begin{table*}[h!t]
    \begin{center}
    \caption{Comparison of SINR and interpolation models at different reconstructed spectral resolutions. PSNR ($\rm {dB})$, SAM, and SSIM are reported. The best results are highlighted in bold.}
    \label{tab1}
    \setlength{\tabcolsep}{6.0pt}{
    \begin{tabular}{ c | c  c  c  c  c  c  c  c  c  c  c  c  c  c }
    \toprule[1.2pt]
    \multirow{2}{*}{Methods}
    & $\times1$ & $\times2$ & $\times4$ & $\times6$ & $\times8$ \\
    & PSNR SAM SSIM & PSNR SAM SSIM & PSNR SAM SSIM & PSNR SAM SSIM & PSNR SAM SSIM\\
    \midrule[0.7pt]
    trilinear ($\times1$) & 28.25, 0.080, 0.961 & 28.07, 0.084, 0.955 & 27.65, 0.084, 0.946 & 27.63, 0.079, 0.948 & 27.52, 0.089, 0.950\\
    
    trilinear ($\times2$) & 29.26, 0.065, 0.965 & 29.39, 0.068, 0.965 & 29.48, 0.069, 0.967 & 29.52, 0.069, 0.968 & 29.62, 0.071, 0.968\\
    
    trilinear ($\times4$)& 30.41, 0.059, 0.966& 30.42, 0.059, 0.966& 30.43, 0.057, 0.967& 30.42, 0.062, 0.964& 30.37, 0.067, 0.960\\
    
    trilinear ($\times6$)& 31.01, 0.052, 0.967& 30.99, 0.051, 0.967& 31.01, 0.051, 0.967& 31.01, 0.051, 0.968& 30.93, 0.052, 0.967\\
    SINR ($\times1$) & \bf{32.24}, \bf{0.048}, \bf{0.969}& \bf{31.37}, \bf{0.046}, \bf{0.969} & \bf{31.40}, \bf{0.046}, \bf{0.969} &\bf{31.43}, \bf{0.046}, \bf{0.970}&\bf{31.39}, \bf{0.046}, \bf{0.970}\\
    \bottomrule[1.2pt]
    \end{tabular}}
    \end{center}
\end{table*}

\begin{table*}[h!t]
    \begin{center}
    \caption{Results of SINR and interpolation with different encoders with different spectral resolutions. PSNR ($\rm {dB}$), SAM, and SSIM are reported. The best results on trilinear and MST are highlighted in bold, respectively.}
    \label{tab2}
    \setlength{\tabcolsep}{6.0pt}{
    \begin{tabular}{ c | c  c  c  c  c  c  c  c  c  c  c  c  c  c }
    \toprule[1.2pt]
    \multirow{2}{*}{Methods}
    & $\times1$ & $\times2$ & $\times4$ &$\times6$ &$\times8$ \\
    & PSNR SAM SSIM& PSNR SAM SSIM& PSNR SAM SSIM& PSNR SAM SSIM& PSNR SAM SSIM\\
    \midrule[0.7pt]
    SSI-trilinear &28.25, 0.069, 0.961 &29.39, 0.068, 0.965 &30.45, 0.057, 0.967 &31.01, 0.051, 0.969 & \bf{31.61}, \rm{0.047}, \bf{0.970}\\
    
    SSI+SINR & \bf{32.24}, \bf{0.048}, \bf{0.969} & \bf{31.37}, \bf{0.046}, \bf{0.969} & \bf{31.40}, \bf{0.046}, \bf{0.969} & \bf{31.41}, \bf{0.046}, \bf{0.970} & 31.39, \bf{0.046}, \bf{0.970}\\
    
    MST+trilinear & 29.35, 0.050, 0.962 & 29.83, 0.052, 0.962 & 30.36, 0.048, 0.967 & 30.43, 0.057, 0.970 & 31.67, 0.064, 0.970\\
    
    MST+SINR & \bf{32.90}, \bf{0.044}, \bf{0.972} & \bf{32.06}, \bf{0.049}, \bf{0.971} & \bf{31.98}, \bf{0.046}, \bf{0.970} & \bf{31.96}, \bf{0.046}, \bf{0.969} & \bf{31.87}, \bf{0.047}, \bf{0.971}\\
    \bottomrule[1.2pt]
    \end{tabular}}
    \end{center}
\end{table*}

\noindent{\bf{Comparison with Fixed-band Interpolation.}} 
This work is the first successful attempt to ensure spectral continuity in the reconstruction process of CASSI. We fairly compare our method with trilinear interpolation using the same reconstruction algorithm. Both methods maintain consistency in training and testing. Notably, our discussed interpolation method differs from simple image super-resolution techniques, as it involves guided interpolation through a neural network with loss constraints. In contrast, our SINR ($\times 1$) performs well across all scales of reconstruction magnification, even with a single trained model.

The results are displayed in Tab.~\ref{tab1}. Using interpolated up-sampling, test performance improves as the training magnification factor increases, but at the expense of longer training times. Each evaluation metric for the interpolation model works best when the test magnification factor matches the training factor. Notably, our SINR achieves a PSNR of $31.37\rm {dB}$ at a testing scale factor of ($\times 2$), surpassing the trilinear ($\times 2$) interpolation's $29.39\rm {dB}$ at the same scale. Similarly, at magnification factors of ($\times 4$) and ($\times 6$), our approach outperforms models specifically trained for these factors using the interpolation method.

These results validate our approach of treating the spectrum as a continuous entity and representing it using an INR function. Our SINR model demonstrates comparable or superior performance to models tailored for individual bands during the reconstruction process. These findings emphasize the effectiveness of our approach in achieving superior reconstruction results and underscore the advantage of considering spectral information as a continuous entity.

\noindent{\bf{Comparison on Different Backbones.}}
In Tab.~\ref{tab2}, we evaluate two HSI reconstruction backbones, namely SSI-ResU-Net~\cite{wang2021new} and MST~\cite{cai2022mask}, paired with both SINR and trilinear interpolation methods. 
Considering that the trilinear method lacks the ability to achieve single-model reconstruction for various spectral scales of HSIs. As a result, the trilinear model results presented in Tab.~\ref{tab2} are obtained by testing a single interpolation model trained for different spectral magnifications. This implies that we train five trilinear models for each backbone configuration.

Our proposed approach significantly outperforms interpolation based on the obtained numerical results. In terms of spectral evaluation, SINR consistently achieves superior SAM values compared to the trilinear method across all scales and encoder backbones, stabilizing around $0.046$. Regarding spatial reconstruction quality, SINR outperforms the interpolation method in terms of PSNR and SSIM across most scales, except for the ($\times 8$) scale. However, even at this scale, SINR remains competitive in terms of SSIM and SAM performance with the trilinear method.

In summary, these quantitative results clearly showcase SINR's ability to effectively reconstruct various spectral bands using a single model. Additionally, the reconstructions achieved by SINR consistently surpass those of the single-scale model, irrespective of the chosen encoder backbone. We attribute this improvement to the incorporation of power spectral features in SINR and its implicit attention mechanisms.

\begin{figure}[h!t]
\centering
\includegraphics[width=\linewidth]{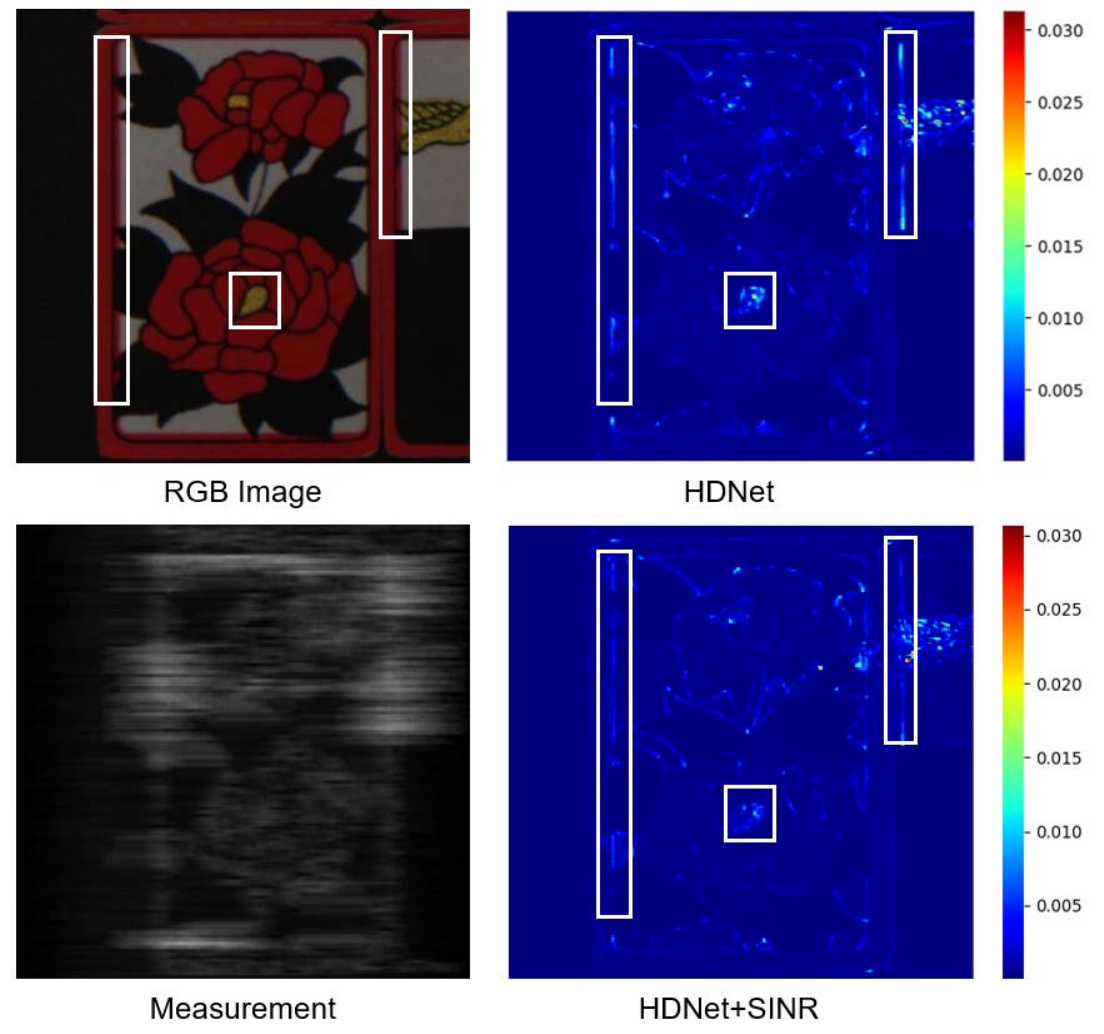}
\caption{Visual error analysis of reconstructed images.
The left column shows the original RGB images and the 2D measurements. The right column shows the error maps (difference between the reconstructed image and the ground truth) without and with SINR, respectively. The model using SINR gives higher fidelity to edge details. }
\label{fig9}
\end{figure}

\begin{table*}
\begin{center}
\caption{Quantitative results of several reconstruct algorithms of 10 scenes on KAIST.}
\label{tab3}
\setlength{\tabcolsep}{3.0pt}{
\begin{tabular}{ l c  c  c  c  c  c  c  c  c  c  c  c  c}
\toprule[1.2pt]
Algorithms& Metrics  & Scene 1 & Scene 2 & Scene 3 & Scene 4 & Scene 5 & Scene 6 & Scene 7 & Scene 8 & Scene 9 & Scene 10 & Average \\
\midrule[0.7pt]
\multirow{9}{*}{MST~\cite{cai2022mask}}
&PSNR$\uparrow$ &34.71& 34.45& 35.32 & 41.50 &31.90 & 33.85 & 32.69 & 31.69 & 34.67 & 31.82 &\textbf{34.26}\\
&SSIM$\uparrow$ &0.930& 0.925& 0.943 & 0.967 &0.933& 0.943 & 0.911 & 0.933 & 0.939 & 0.926 &0.935\\
&SAM$\downarrow$
&0.131&0.191&0.123&0.179&0.131&0.198&0.122 &0.236&0.149&0.199&0.166\\
&VSNR$\uparrow$
&30.25& 29.72& 32.58& 39.93& 28.11& 28.56& 29.03& 27.11& 30.64& 25.99 & 30.20 \\
&WSNR$\uparrow$
&38.12& 39.20& 34.73& 38.02& 37.91&37.30&38.90&38.56& 37.26& 38.14 &37.82 \\
&NQM$\uparrow$&16.68& 17.38& 19.27& 16.53& 17.60&20.08& 16.94& 17.63& 17.76& 14.69&17.46
                \\
&IFC$\uparrow$&2.207 &2.406	&1.604	&0.816	&1.959	&1.747	&1.762&	1.825	&1.619&	1.849&1.779
                  \\
&UQI$\uparrow$&0.979&0.979&0.978& 0.946&0.983&0.987& 0.975&0.976& 0.984& 0.974&0.976                                                 \\
&VIF$\uparrow$&0.841&0.829&0.927&0.957&0.851&0.879&0.859&0.836&0.931&0.731&0.864
                   \\
\hline
\multirow{9}{*}{MST+SINR}
&PSNR$\uparrow$&34.50 & 34.26& 34.95 & 42.58& 31.72  & 33.59 & 32.86 & 31.74 & 34.22 &31.46&34.19\\
&SSIM$\uparrow$ & 0.933 & 0.930 & 0.948& 0.981 & 0.937 & 0.949 &0.925 & 0.944& 0.943 & 0.929 &\textbf{0.942}\\
&SAM$\downarrow$ &0.130 & 0.149 &0.105 &0.138 &0.099& 0.139&0.111&0.155& 0.117&0.134&\textbf{0.128}\\
&VSNR$\uparrow$ &30.77& 30.10& 32.65&40.39&28.44& 29.29&29.14& 27.41& 31.07& 26.42 & \textbf{30.57} \\
&WSNR$\uparrow$ &38.62& 39.38&34.95& 38.05& 38.12& 37.92& 38.95&39.06&37.28& 38.73 &\textbf{38.10} \\
&NQM$\uparrow$&16.93& 17.85& 19.29& 16.88&18.16&20.09& 17.09&17.79&17.91&15.15&\textbf{17.71}
                   \\
&IFC$\uparrow$&2.246&2.415	&1,702	&0.824	&1.964	&1.772	&1.793	&1.838&	1.667&1.888&\textbf{1.811}
                  \\
&UQI$\uparrow$&0.981&0.980&0.981&0.959& 0.985& 0.988& 0.979& 0.979& 0.985&0.977&\textbf{0.979}
             \\
&VIF$\uparrow$&0.866&0.859&0.949&0.958& 0.858&0.926& 0.881& 0.869&0.967&0.765&\textbf{0.889}
                   \\
\hline
\multirow{9}{*}{HDNet~\cite{hu2022hdnet}}
&PSNR$\uparrow$& 34.95 & 32.52 & 34.52 & 43.00 & 32.49 & 35.96 & 29.18 & 34.00 & 34.56 & 32.22 & 34.34\\
&SSIM$\uparrow$ &0.948 & 0.953 &0.957 &0.981& 0.956 & 0.964 & 0.937&0.961 & 0.957 &0.950 &0.957\\
&SAM$\downarrow$ &0.132& 0.151&0.104& 0.138& 0.098& 0.145& 0.115& 0.166&0.124& 0.147&0.132\\
&VSNR$\uparrow$  &31.21& 31.23& 33.53& 40.55& 29.17& 29.87& 29.87& 27.89& 31.62& 27.53  &31.25 \\
&WSNR$\uparrow$&36.94& 36.64& 32.86& 36.57& 36.60& 35.59& 36.85& 36.88& 34.82& 37.04 &36.08  \\
&NQM$\uparrow$&17.41&18.69&20.22&16.31&18.69&21.09&17.99&18.32&18.52&18.28&18.55
                   \\
&IFC$\uparrow$&2.413&2.735&	1.748&	0.857&	2.177&	1.893&	1.999&	1.945&1.807&	2.047&1.962
                             \\
&UQI$\uparrow$&0.982&0.984&0.985&0.954&0.986&0.99&0.981&0.981& 0.985& 0.979&0.980
                         \\

&VIF$\uparrow$&0.875&0.852&0.924&0.885&0.864&0.898&0.874& 0.877&0.932&0.799&0.877                                                                        \\
\hline

\multirow{9}{*}{HDNet+SINR}
&PSNR$\uparrow$ &35.08 & 32.85& 35.06 & 43.21 & 32.69 & 36.01& 29.31& 34.09& 35.06&32.16&\textbf{34.55}\\
&SSIM$\uparrow$ &0.949 & 0.956& 0.963& 0.985&0.958&0.966& 0.942&0.963&0.959&0.950&\textbf{0.959}\\
&SAM$\downarrow$ &0.117&0.126&0.082&0.124& 0.089&0.126&0.098& 0.143&0.105&0.129&\textbf{0.114}\\
&VSNR$\uparrow$  &31.80& 32.19& 34.86& 41.76& 29.75& 30.77& 30.75& 29.05& 33.37& 27.61  & \textbf{32.19}  \\
&WSNR$\uparrow$ & 37.57& 37.58& 34.33& 37.93& 37.38& 36.71& 37.76& 38.37& 36.48& 37.37 & \textbf{37.15} \\
&NQM$\uparrow$&18.07& 20.18& 21.83& 18.41& 19.69& 22.48& 19.33& 19.72& 20.40& 15.97&\textbf{19.60}
                     \\
&IFC$\uparrow$&2.638&3.105&1.962&0.989&2.421&2.151&2.176&2.268&2.036&2.249&\textbf{2.199}

                 \\
&UQI$\uparrow$&0.985& 0.988&0.99& 0.97& 0.989&0.992&0.986&0.985&0.99&0.981&\textbf{0.986}
                  \\

&VIF$\uparrow$&0.882&0.874&0.945&0.917&0.871& 0.917& 0.884&0.897&0.973&0.800&\textbf{0.896}
                  \\

\bottomrule[1.2pt]
\end{tabular}}
\end{center}
\end{table*}

\begin{table*}

\begin{center}
\caption{Average quantitative results of five reconstruct algorithms on Harvard dataset.}
\label{Harvard}
\setlength{\tabcolsep}{1.0pt}{
\begin{tabular}{ l c  c  c  c  c  c  c  c  c  c }
\toprule[1.2pt]
Metrics& TSA-Net~\cite{wang2021tsa} & TSA-Net+SINR & MST~\cite{cai2022mask} & MST+SINR  &HDNet~\cite{hu2022hdnet}  &HDNet+SINR  &CST~\cite{cai2022coarse}  &CST+SINR  &SSI~\cite{wang2008new} &SSI+SINR\\
\midrule[0.7pt]
PSNR$\uparrow$  &34.13  &\textbf{34.67}  &35.72  &\textbf{35.98}  &35.83  &\textbf{36.10}  &35.41  &\textbf{35.77}  &34.87  &\textbf{35.40}  \\
SSIM$\uparrow$  &0.820  &\textbf{0.834}  &0.853  &\textbf{0.864}  &0.855  &\textbf{0.865}  &0.853  &\textbf{0.860}  &0.834  &\textbf{0.857}  \\
SAM$\downarrow$  &0.090  &\textbf{0.085}  &0.080  &\textbf{0.073}  &0.084  &\textbf{0.078}  &0.086  &\textbf{0.077}  & 0.099&\textbf{0.082}  \\
VSNR$\uparrow$  &29.96  &\textbf{30.19}  &30.95  &\textbf{31.28}  &31.11  &\textbf{31.46}  &30.79  &\textbf{31.32}  &30.27  &\textbf{31.02}  \\
WSNR$\uparrow$  &32.72  &\textbf{33.18}  &31.62  &\textbf{32.31}  &31.65  &\textbf{31.95}  &32.19  &\textbf{32.73}  &32.21  &\textbf{32.69}  \\
NQM$\uparrow$&6.25&\textbf{6.77}&7.53&\textbf{8.04}&7.57&\textbf{8.45}&7.32&\textbf{7.70}&6.43&\textbf{7.36}
                 \\
IFC$\uparrow$&0.512&\textbf{0.580}&0.795&\textbf{0.807}&0.758&\textbf{0.842}&0.708&\textbf{0.779}&0.597&\textbf{0.696}
                       \\
UQI$\uparrow$&0.769&\textbf{0.793}&0.817&\textbf{0.839}&0.820&\textbf{0.847}&0.812&\textbf{0.829}&0.786&\textbf{0.811}
                           \\
VIF$\uparrow$&0.650&\textbf{0.677}&0.711&\textbf{0.720}&0.743&\textbf{0.787}&0.687&\textbf{0.709}&0.681&\textbf{0.712}\\

\bottomrule[1.2pt]
\end{tabular}}
\end{center}
\end{table*}

\noindent{\bf{Effect of SINR on CAVE and KAIST.}} 
Based on the reconstructed CASSI reconstruction algorithms, SINR serves as a plug-and-play module. We integrate SINR with advanced HSI reconstruction methods, including MST~\cite{cai2022mask} and HDNet~\cite{hu2022hdnet}, to assess the impact of incorporating SINR.
Our training is based on the KAIST dataset, and validation is conducted using the CAVE dataset. Similar to earlier experiments, we interpolate 28 bands as input, with all 28 bands serving as ground truth for training. The comparison focuses on 1:1 $(\times 1)$ reconstruction, as the existing algorithms lack spectral super-resolution capability.

From Tab.~\ref{tab3}, it can be observed that our method not only enhances pixel fidelity, as indicated by higher PSNR values, but also improves visual quality by reducing perceived noise, as evidenced by higher VSNR and VIF values. The superior performance of MST+SINR in terms of spectral similarity, with a margin of $0.038$ which highlight the high fidelity of the spectral reconstruction achieved by SINR. The increased NQM and IFC metrics further emphasize the noise reduction effect achieved by SINR. The incorporation of SINR also enhances the performance of the WSNR metric, underscoring the significance of our method in considering the importance of different frequency components and overall signal quality. In conclusion, the consistency of these metric trends underscores the robustness and effectiveness of our method in various aspects of image quality.

\noindent{\bf{Effect of SINR on Harvad.}}
We further validated SINR's reconstruction performance on the Harvard dataset, comparing it with five methods: TSA-Net~\cite{wang2021tsa}, MST~\cite{cai2022mask}, HDNet~\cite{hu2022hdnet}, CST~\cite{cai2022coarse}, and SSI~\cite{wang2008new}. The experimental setup matches CAVE and KAIST datasets. Results for ten test images and their averages are in Tab.~\ref{Harvard}.

Upon integrating SINR into the TSA-Net algorithm, there is a significant enhancement across all metrics. Notably, in comparison to the standalone TSA-Net algorithm, the introduction of SINR leads to a substantial increase in the PSNR value, elevating it from $34.13\rm{dB}$ to $34.67\rm{dB}$. Furthermore, the SSIM value demonstrates noticeable improvement, rising from $0.820$ to $0.834$, indicating an enhancement in structural similarity. Moreover, the SAM value experiences a considerable reduction, dropping from $0.090$ to $0.085$, suggesting a reduction in spectral angle mismatches. The elevated VSNR and WSNR values underscore the positive impact of SINR on enhancing visual perception, noise reduction, and signal quality across different frequency components.

Furthermore, the increase in NQM value highlights the method's capacity to mitigate noise levels within reconstructed images, progressing from $17.46$ to $17.71$. Similarly, the higher IFC value substantiates the method's improved preservation of high-frequency details, rising from $1.779$ to $1.881$. UQI's comprehensive assessment of spatial and spectral features, along with VIF's preservation of perceptually relevant information, are also amplified through the integration of SINR.

Moreover, similar positive trends are observed when SINR is integrated into the MST, HDNet, CST, and SSI algorithms.
The consistent and positive trends across all nine metrics collectively validate the robustness of our proposed method in enhancing visual quality and noise reduction in reconstructed images. These improvements hold steadfast across various datasets, showcasing elevated spectral fidelity, robustness, and versatility brought about by SINR.

\subsection{Qualitative Results}

\noindent{\bf{Reconstruction on ICVL.}} 
We present visual examples of the ($\times 2$) and ($\times 4$) tasks using the ICVL dataset. The CASSI system compresses 31 input image channels, which our model reconstructs into 62 and 186 bands, respectively. Test images are of size $60\times60$, displayed as $120\times120$ by stitching four images together. In Fig.~\ref{fig4}, we showcase five selected bands for comparison. Notably, our SINR-generated images exhibit clearer textures and fewer artifacts than those produced by interpolation.

Four points are randomly selected (marked in RGB images) to compute curve correlations between both methods and reference values across the spectral range of 450$\rm {nm}$ to 650$\rm {nm}$. The curves reveal that trilinear interpolation results in consistently smooth spectral curves with fewer details. In contrast, SINR accurately captures and models spectral curve intricacies. Additionally, for the four marked points, SINR demonstrates higher spectral correlation values: $0.9381$, $0.9659$, $0.9563$, and $0.9552$, compared to trilinear interpolation values of $0.9127$, $0.9484$, $0.8706$, and $0.8384$ respectively. These outcomes highlight SINR's capacity for improved spectral-wise attention and FCE's adeptness in capturing and modeling spectral details, resulting in higher correlation and more intricate spectral curves.

\noindent{\bf{Reconstruction on CAVE and KAIST.}}
Fig.~\ref{fig5}, we depict the reconstruction outcomes of two methods on KAIST, employing the same backbone.  Since the dataset contains only 28 frequency bands, evaluation of the amplification effect and performance of the interpolating function is only supported on a ($\times 1$) scale. To underscore SINR's spectrum modeling prowess, we focus on spatial reconstruction and illustrate the spectral density curves at two spatial points, denoted as "a" and "b" in Fig.~\ref{fig5}.
Comparing the spectral density curves, SINR-generated curves more closely align with the ground truth spectrum, particularly in capturing intricate spectral structures. This robustly underscores SINR's potent spectral modeling ability.

\noindent{\bf{Reconstruction on Harvard.}}
Fig.~\ref{harvad} showcases the outcomes of our Harvard dataset reconstruction. We present four distinct wavelength images extracted from the reconstructed HSIs. A notable enhancement in overall reconstruction quality is evident upon comparing images with and without SINR integration. The images exhibit improved clarity and reduced edge blurring. Upon closer inspection in the bottom right corner, a significant reduction in image blurring is observed, along with enhanced representation of fine structures due to SINR's inclusion.

\begin{figure}[htbp]
\centering
\includegraphics[width=\linewidth]{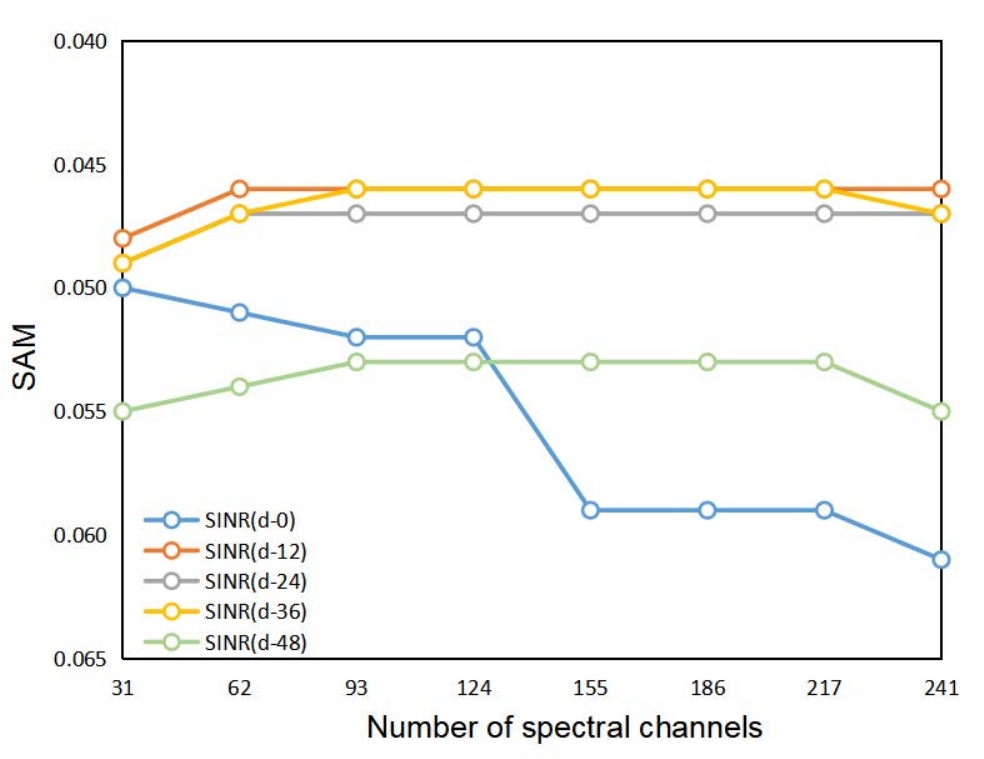}
  \caption{Performance on SAM for different mapping dimensions. SINR(d-0)/(d-12)/(d-24)/(d-36)/(d-48) corresponds to the SINR model with mapping dimensions 0/12/24/36/48.}
\label{fig8}
\end{figure}

Furthermore, the spectral density curves on the right side demonstrate the superiority of SINR. The overall trend and spectral details of the SINR-reconstructed curves closely align with the ground truth, surpassing a correlation coefficient of $0.996$. This indicates that the SINR reconstruction yields spectra that are not only consistent in general trends but also capture spectral nuances more accurately.

\noindent{\bf{Error Map of Reconstructed Images.}}
Fig.~\ref{fig9} illustrates the error map contrasting reconstructed images and the ground truth. The spectral channel of the reconstructed image, representing a ($\times 4$) scale, is 124. Brighter shades denote higher errors, while bluer shades indicate lower errors. Notably, the SINR-enhanced model displays reduced errors in the edge regions, closely aligning with the ground truth. These outcomes affirm that SINR yields enhanced accuracy and sharper delineation in complex structural areas, attributable to its emphasis on high-frequency information.

\begin{table*}[t]
\centering
\setlength{\tabcolsep}{7.5pt}
\caption{Ablations on proposed components. The PSNR ($\rm {dB}$) and SAM results are the average value of all test images on ICVL.}
\begin{tabular}{ccc cc cc cc cc cc}
\toprule[1.2pt]
 \multirow{2}{*}{SWA} & \multirow{2}{*}{FCE} & \multirow{2}{*}{SF} & \multicolumn{2}{c}{$\times$1} & \multicolumn{2}{c}{$\times$2} & \multicolumn{2}{c}{$\times$4} &\multicolumn{2}{c}{$\times$6}& \multicolumn{2}{c}{$\times$8} \\
 &  &  & PSNR  & SAM & PSNR& SAM & PSNR & SAM & PSNR & SAM & PSNR & SAM\\ 
\midrule[0.7pt]
\color{red} \ding{55}& \color{red} \ding{55} & \color{red} \ding{55} &28.32&0.078  &27.55& 0.078 &27.26&0.075 &27.32&0.075 &26.84&0.076\\
\color{red} \ding{55}  & \color{red} \ding{55} &  \color{green} \ding{51} &28.65&0.075  &28.33& 0.074 &27.71&0.075 &27.49&0.073 &27.18&0.073\\
 \color{red} \ding{55} &  \color{green} \ding{51}& \color{red} \ding{55} &29.46&0.068  &29.35& 0.066 &28.74&0.067 &28.03&0.068 &27.62&0.067\\
 \color{green} \ding{51} &\color{red} \ding{55}  & \color{red} \ding{55} & 30.06 & 0.054 & 29.92 & 0.051 & 29.13 & 0.052 &28.07 & 0.053 & 27.87 &0.053\\

 \color{red} \ding{55}&  \color{green} \ding{51} &\color{green} \ding{51}& 30.42& 0.062 & 30.50& 0.060 & 30.50&0.060 &30.47& 0.061& 30.31& 0.064\\
\color{green} \ding{51} & \color{red} \ding{55}&  \color{green} \ding{51} & 31.13 & 0.050 & 30.04 & 0.051 & 29.21 & 0.052 &28.42 & 0.051 & 28.20 &0.051\\
  \color{green} \ding{51} & \color{green} \ding{51} & \color{red} \ding{55}& 31.46& 0.049& 31.66& 0.047& 31.68& 0.047 &30.05& 0.055& 30.05& 0.055\\
 \color{green} \ding{51} & \color{green} \ding{51} &  \color{green} \ding{51} & \textbf{32.24} & \textbf{0.048} & \textbf{31.37}& \textbf{0.046} & \textbf{31.40}& \textbf{0.046 }&\textbf{31.41}& \textbf{0.046} & \textbf{31.39}& \textbf{0.046} \\
\bottomrule[1.2pt]
\end{tabular}
\label{tab4}
\end{table*}

\begin{table*}[h!t]
    \begin{center}
    \caption{Performance on different mapping dimensions in FCE. SINR(d-0)/(d-12)/(d-24)/(d-36)/(d-48) corresponds to the SINR model with mapping dimensions 0/12/24/36/48. }
    \label{tab5}
    \setlength{\tabcolsep}{13.0pt}{
    \begin{tabular}{c | c  c  c  c c}
    \toprule[1.2pt]
    \multirow{2}{*}{Mapping Dimension}
    & $\times1$ & $\times2$ & $\times4$ & $\times6$ & $\times8$\\
    & PSNR SAM  & PSNR SAM & PSNR SAM  &PSNR SAM  &PSNR SAM \\
    \midrule[0.7pt]
    SINR(d-0) & 31.13, 0.050 & 30.04, 0.051 & 29.21, 0.052 & 28.42, 0.059 & 28.20, 0.061 \\

    SINR(d-12) & 31.24, 0.048 & 31.37, 0.046 & 31.40, 0.046  & 31.41, 0.046 & 31.39, 0.046 \\

    SINR(d-24) & 31.07, 0.049 & 31.18, 0.047 & 31.22, 0.047 & 31.22, 0.047 & 31.19, 0.047\\
    
    SINR(d-36) & 31.31, 0.049 & 31.43, 0.047& 31.48, 0.047 & 31.49, 0.048 & 31.49, 0.048 \\
    
    SINR(d-48) & 31.17, 0.055 & 31.29, 0.054& 31.34, 0.053 & 31.34, 0.053 & 31.25, 0.055\\
    \bottomrule[1.2pt]
    \end{tabular}}
    \end{center}
\end{table*}

\section{Ablation Study}
\label{sec:ablation}
Extensive ablation studies are conducted on the ICVL dataset to validate our design choices and parameter settings. We first performed ablation experiments for various cases for the proposed module, and the results are presented in Tab.~\ref{tab4}.

\noindent {\bf{Effect of Spectral-wise Attention.}} 
Spectral-wise attention is a critical component of SINR by treating each spectral channel as a token. This approach enables SINR to gather non-local feature information from neighboring bands, amplifying channel-specific attributes. The baseline model solely combines encoder output features with original coordinates and undertakes INR via MLP decoding, excluding SWA, FCE, or SF modules. In Tab.~\ref{tab4}, upon integrating SWA, PSNR elevates from $28.32\rm {dB}$ to $30.06\rm {dB}$, and SAM enhances by $0.024$. These outcomes underscore that the spectral emphasis introduced by SINR empowers the network to ensure precise spectrum reconstruction, markedly enhancing the quality of reconstructed images.

\noindent{\bf{Effect of Fourier coordinate encoder.}}
The Fourier coordinate encoder aims to enhance the model by comprehensively modeling high-frequency information, thus recovering fine-grained spectral details. As demonstrated in Tab.~\ref{tab4}, integrating FCE with the baseline model results in a substantial increase of $1.14\rm {dB}$ for PSNR while reducing SAM from $0.078$ to $0.068$. The integration of FCE brings substantial enhancements in both PSNR and SAM metrics, underscoring the pivotal role of mapping coordinates into higher-dimensional spaces. Neglecting the encoding of high-dimensional coordinates in favor of low-frequency information is suboptimal. By enabling the network to retain richer high-frequency details, FCE aids in reconstructing fine HSI structures.

\noindent {\bf{Effect of Scale Factor.}} 
The scale factor incorporates a global amplification factor into the model, which contributes to scale-related analysis. The presence of SF in Tab.~\ref{tab4} demonstrates improvements in both PSNR and SAM compared to the model without SF. This highlights the essential nature of this design in enhancing the model's reconstruction performance.

\noindent {\bf{Interaction between modules.}} 
A comparison in Tab.~\ref{tab4} reveals that incorporating two or more modules together with the baseline yields greater improvements in PSNR and SAM than adding each module individually. For instance, when SF and FCE are added separately, PSNR improves by $0.33\rm {dB}$ and $1.14\rm {dB}$, with SAM improvements of $0.003$ and $0.01$ respectively. However, when both FCE and SF are integrated simultaneously, PSNR improves by $2.10\rm {dB}$ and SAM improves by $0.016$. These findings suggest that the proposed modules exhibit a synergistic interaction that goes beyond a simple additive effect, leading to enhanced reconstruction performance of the model.

\noindent {\bf{Impact of Fourier Coordinate Mapping.}} 
We study the impact of Fourier coordinate mapping dimensions on model performance across different magnifications. Five dimensions, namely 0, 12, 24, 36, and 48, are evaluated for reconstruction quality. As seen in Tab. \ref{tab5}, varying mapping dimensions are tested at different levels of magnification. Notably, at a mapping dimension of 36, the highest PSNR value of $31.49 \, \rm{dB}$ is achieved, outperforming other dimensions. Both PSNR and SAM metrics exhibit peaks rather than a consistent linear relationship with increasing mapping dimensions.

Fig. \ref{fig8} further illustrates that Fourier coordinate mapping effectively mitigates performance degradation with an increasing number of reconstructed spectral channels. Importantly, a higher mapping dimension does not necessarily correspond to a more accurate reconstructed spectrum. The best spectral reconstruction accuracy occurs at a mapping dimension of 12, while the largest SAM value is observed at a mapping dimension of 48. 

\section{Discussion}
By incorporating the continuous implicit representation of SINR, we enable the training of a single model for reconstructing HSIs across various spectral band numbers. This eliminates the need for retraining and saves time. While SINR can enhance other reconstruction algorithms, many existing methods are computationally intensive. Moreover, training with SINR requires considerable time. Our future work aims to develop a lightweight reconstruction algorithm that, when combined with SINR, can efficiently reconstruct arbitrary spectral bands.

Furthermore, we plan to extend SINR's capabilities beyond spectral band reconstruction to achieve spatially continuous amplification. This advancement will lead to a lightweight, dual-domain spatial-spectral reconstruction framework, enhancing the flexibility of CASSI reconstruction for HSIs. These research directions will advance HSI reconstruction, making it more accessible, efficient, and adaptable to diverse applications.

\section{Conclusion}
This study introduces a novel implicit neural representation for HSI reconstruction, enabling spectral super-resolution at arbitrary and controllable ratios. It marks the first comprehensive fusion of implicit neural representation and CASSI reconstruction. The model adapts to spectral channel relationships and multi-scale coordinates, treating HSI as a continuous function of spectral coordinates. Additionally, it incorporates tailored spectral-wise attention and Fourier coordinate encoder to enhance performance. Both qualitative and quantitative evaluations demonstrate that our SINR yields visually appealing and high-quality reconstructed images, maintaining its effectiveness across various scales even when trained on a fixed scale.


\bibliographystyle{IEEEtran}
\bibliography{egbib}

\end{document}